\documentclass[12pt]{article}
\usepackage[dvipdfmx]{graphicx}
\usepackage{amsfonts,amsmath,amssymb,epsf}
\usepackage{amsmath,braket}
\usepackage{hyperref,color}
\usepackage[toc,page]{appendix}
\usepackage{tikz}
\hypersetup{
    bookmarks=true,         % show bookmarks bar?
    unicode=false,          % non-Latin characters in Acrobat’s bookmarks
    pdftoolbar=true,        % show Acrobat’s toolbar?
    pdfmenubar=true,        % show Acrobat’s menu?
    pdffitwindow=false,     % window fit to page when opened
    pdfstartview={FitH},    % fits the width of the page to the window
    pdfnewwindow=true,      % links in new PDF window
    colorlinks=true,       % false: boxed links; true: colored links
    linkcolor=blue,          % color of internal links (change box color with linkbordercolor)
    citecolor=blue,        % color of links to bibliography
    filecolor=blue,      % color of file links
    urlcolor=blue           % color of external links
}
\usepackage{hyperref}

\usepackage[T1]{fontenc}
\usepackage[latin9]{inputenc}

\usepackage{braket}
\usepackage{amsmath}
\usepackage{amssymb}
\usepackage{color}

\makeatletter
\numberwithin{figure}{section}
\numberwithin{equation}{section}
\numberwithin{table}{section}

\newcommand{\be}{\begin{equation}}
\newcommand{\ba}{\begin{eqnarray}}
\newcommand{\ea}{\end{eqnarray}}
\newcommand{\ee}{\end{equation}}

\newcommand{\bea}{\begin{eqnarray}}
\newcommand{\eea}{\end{eqnarray}}
\newcommand{\bes}{\begin{equation*}}
\newcommand{\beas}{\begin{eqnarray*}}
\newcommand{\eeas}{\end{eqnarray*}}
\newcommand{\bas}{\begin{array*}}
\newcommand{\eas}{\end{array*}}
\newcommand{\ees}{\end{equation*}}
\newcommand{\nn}{\nonumber}
\newcommand{\p}{\partial}
\newcommand{\ep}{\epsilon}

\makeatother
\usepackage{comment}
\usepackage{babel}

\begin{document}

\begin{titlepage}
\thispagestyle{empty}

\begin{flushright}
KUNS-2982
\\
YITP-23-101
\\
\end{flushright}

\bigskip

\begin{center}
\noindent{\Large \bf End of the World Branes from Dimensional Reduction }\\
\vspace{2cm}

Shigeki Sugimoto$^{\dagger,*,\star}$ and 
Yu-ki Suzuki$^*$
\vspace{1cm}\\

{\it $^\dagger$ Department of Physics, Kyoto University, Kyoto 606-8502, Japan}

{\it $^*$Center for Gravitational Physics and Quantum Information,\\
Yukawa Institute for Theoretical Physics, Kyoto University, \\
Kitashirakawa Oiwakecho, Sakyo-ku, Kyoto 606-8502, Japan}

{\it $^\star$Kavli Institute for the Physics and Mathematics of the Universe, University of Tokyo, Kashiwano-ha, Kashiwa, Chiba 277-8582, Japan}\\

%\vskip 2em
\end{center}

\begin{abstract}
We consider dimensional reduction of cigar geometries which are obtained by a Wick rotation of black hole solutions. Originally the cigar geometry is smooth around the tip, but after the dimensional reduction along the Euclidean time direction, there appears an end-of-the-world brane (ETW brane). We derive the tension of the brane by two methods: bulk equations of motion and boundary equations of motion.

In particular, for AdS$_7$-soliton $\times$S$^4$ and AdS$_4$-soliton $\times$S$^7$ backgrounds in M-theory, we find that the tension of the emerging ETW branes behaves as $t(\Phi)\sim e^{-3\Phi}$ in the string frame. This indicates the existence of such ETW branes in the strongly coupled regime of type 0A string theory.

%As a consistency check, we also apply our methods to a system with an O8-plane and D8-branes, and reproduce the known value of the tension in superstring theory.

\end{abstract}

\end{titlepage}

\newpage

\tableofcontents
\newpage
\section{Introduction and Summary}
The end-of-the-world brane (ETW brane) is a codimension one brane at which the space-time ends. It appears in various theories with gravity defined in a space-time with boundaries. It could couple with gravity as well as other fields, and some of the degrees of freedom may be localized on it. In string theory, such an object has been studied for long time since it can be obtained as an orientifold-plane or a fixed plane of a ${\bf Z}_2$ orbifold and played important roles in understanding various dualities in string theory. A well-known example is an M-theory lift of the $E_8\times E_8$ Heterotic string theory, which is obtained by compactifying the 11-dimensional space-time on an interval $S^1/{\bf Z}_2$ \cite{Horava:1996ma}. The ETW branes are located at the boundary of the interval, and each of them carries an $E_8$ gauge theory on it. This system is related by S and T-dualities to type IIA string theory with two orientifold 8-planes (O8-planes)\footnote{See e.g. \cite{Polchinski:1996fm} for a review}, which is also called type I' or type IA string theory. This O8-plane is also an example of the ETW brane realized in string theory.

More recently, the ETW branes are studied in the context of AdS/BCFT correspondence\cite{Karch:2000gx,Takayanagi:2011zk,Fujita:2011fp}, which is a holographic duality between conformal field theory (CFT) with boundaries called boundary confomral field theory (BCFT)\footnote{See \cite{Cardy:2004hm} for a nice review of BCFT.} and theory of gravity in asymptotically anti-de Sitter (AdS) space-time with ETW branes. It has been applied to many physical problems such as the Kondo effect \cite{Erdmenger:2013dpa}, the information paradox \cite{Penington:2019npb,Almheiri:2019hni,Penington:2019kki,Almheiri:2019qdq} and the measurement-induced phase transition  \cite{Kanda:2023zse,Kanda:2023jyi}, and provided holographic explanations of various phenomena including relation to quantum information theory \cite{Suzuki:2022tan}. In most of these works, however, the ETW branes are introduced by hand in bottom up models of the AdS/BCFT correspondence. It would be nice to have top down realizations of the ETW branes in string theory (see \cite{Uhlemann:2021nhu,VanRaamsdonk:2021duo,Karch:2022rvr,Harvey:2023pdv} for recent attempts).

In this paper, we pursue novel examples of the ETW brane. Usually, in the bottom up models, the tension of the ETW brane as well as other interaction to the bulk fields are assumed and dynamics of the brane is determined by solving the equations of motion (EOM). Unlike such an approach, we start with a smooth geometry without boundaries (except for the asymptotic boundary), and find that the ETW brane emerges through dimensional reduction. The tension of the ETW brane is determined from consistency, as explained below in detail.

\begin{figure}[h]
    \begin{center}
        \includegraphics[width=120mm]{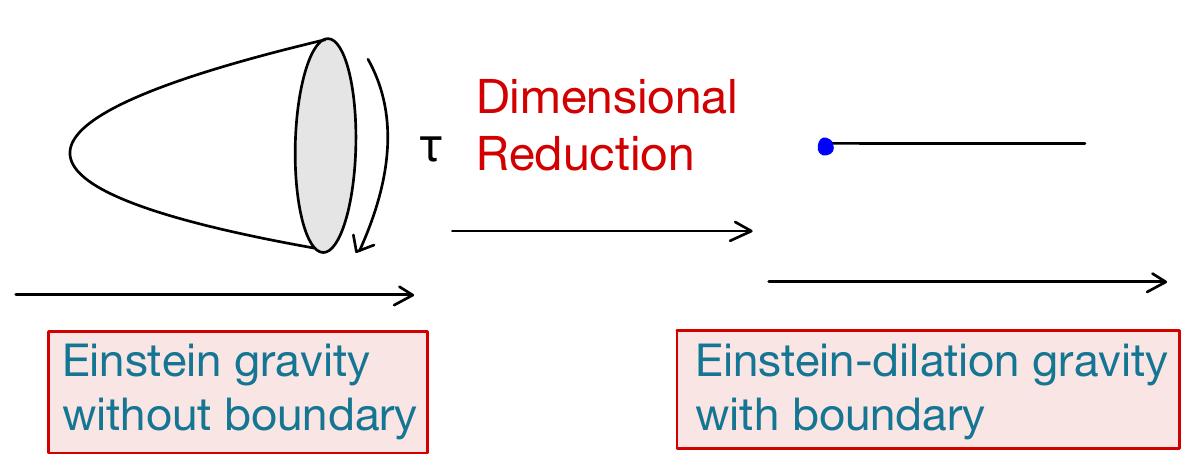}
        \vspace*{-10pt}
    \end{center}
    \caption{Dimensional reduction of the cigar geometry. The left figure represents the cigar geometry. This geometry is smooth everywhere and there are no boundaries (except for the asymptotic boundary). The right figure describes the geometry after dimensional reduction along $\tau$. Accordingly, there emerges a physical boundary (ETW brane), which is represented by the blue dot. }
    \label{dimred}
\end{figure}

To be more concrete, consider a cigar geometry obtained by the Wick rotation of a static black hole (brane) solution. There is a $U(1)$ isometry acting as a constant shift of the Euclidean time coordinate $\tau$ and an orbit of the $U(1)$ action form an $S^1$ whose radius shrinks to a zero size at the horizon. A two-dimensional surface parametrized by $\tau$ and the radial coordinate $r$ looks like a cigar as depicted in the left panel of figure \ref{dimred}. The tip of the cigar corresponds to the horizon in the original black hole (brane) geometry. The AdS-soliton geometry, obtained by the Wick rotation of the AdS-Schwarzschild solution, and Wick rotated black D-brane solutions are important examples of such kind, which are used
to obtain the holographic dual descriptions of non-supersymmetric gauge theories.\footnote{See \cite{Aharony:1999ti} for a review.} Before the dimensional reduction, the cigar geometry is smooth everywhere.
However, the geometry after the dimensional reduction along the Euclidean time direction has a boundary (ETW brane) at the tip of the cigar as depicted in the right panel of figure \ref{dimred}, (see \cite{Friedrich:2023tid} for a similar setup). Taking into account that through dimensional reduction, a part of metric component give rise to the dilaton, our statement is summarized as:

\vspace{4mm}
\textit{Einstein gravity (with some matter fields) on the $(d+1)$-dimensional cigar geometry induces the Einstein-dilaton gravity (with some matter fields) with a  $(d-1)$-dimensional ETW brane in $d$-dimensions upon dimensional reduction along the $\tau$ direction.}
\vspace{4mm}

Note that the ETW brane obtained in this way is a co-dimension one brane, which couples with the dilaton field as well as gravity. Our main focus is on calculating the dilaton dependent tension of the ETW brane via two apparently different methods: one of them makes use of the bulk EOM and the other uses the boundary EOM.
For the method using bulk EOM, we first note that the metric and the dilaton field satisfy the EOM for the Einstein-dilaton theory obtained by the dimensional reduction, because the background cigar geometry we used is a solution of the Einstein equation in $(d+1)$ dimensions. Then, we introduce a $\mathbf{Z}_2$ orbifold as a guide to evaluate the tension of the ETW brane so that it can be a source of gravity and the dilaton field consistent with the bulk solution.\footnote{This method was recently used in \cite{Raucci:2022jgw} for the study of codimension one branes in string theory.} For the other method, the tension of the ETW brane is fixed by solving the boundary EOM as in \cite{Israel:1966rt,Takayanagi:2011zk,Fujita:2011fp}. We can see that the energy-momentum tensor on the brane contributes to the boundary EOM, enabling us to evaluate the tension as well. 
As a consistency check, we also apply these methods to a system with an O8-plane and D8 branes in type IIA superstring theory, and show that the known values of the tensions are reproduced.

For example, we carry out the calculation in AdS-soliton backgrounds with or without internal geometies (spheres). For the purely $d+1$-dimensional AdS-soliton background, the tension of the resulting ETW brane is proportional to $t(\Phi)=-de^{-\sqrt{\frac{d-1}{2(d-2)}}\Phi}$, where $\Phi$ is the dilaton field. Note that only for $d=3$ the coefficient of $\Phi$ in the exponent becomes an integer. As an application to string theory, we also consider the AdS$_7\times$S$^4$ and the AdS$_4\times$S$^7$ soliton solutions in M-theory and take the dimensional reduction along the $\tau$ direction. Since the fermions of the system obey anti-periodic boundary condition along the Euclidean time direction, the supersymmetry is completely lost and the theory is conjectured to be equivalent to type 0A string theory \cite{Bergman:1999km}. Therefore, the ETW branes obtained by the dimensional reduction are expected to exist in type 0A string theory. In both backgrounds, it turns out that the tension of the ETW branes behaves as $t(\Phi)\sim e^{-3\Phi}$ in the string frame, which does not appear in the perturbative string coupling expansion series. However, this is not a contradiction because the supergravity background in 10 dimensions is singular at the ETW brane and the perturbative calcultions in string theory cannot be trusted. It would be interesting to have further evidence of the existence of such ETW branes.

This paper is organized as follows.
In section \ref{sec2}, first we fomulate the general action after the dimensional reduction and derive the EOM. Then, we explain how to fix the tension from the bulk EOM via $\mathbf{Z}_2$ orbifold and boundary EOM. We show that these two methods give the same answer.
In section \ref{sec3}, we present our calculations in various examples. We firstly consider AdS$_{d+1}$ soliton geometry and then also extend to AdS$_4\times$S$^7$ and AdS$_7\times$S$^4$ geometries. In particular, we obtain a peculiar dilaton dependence of the ETW branes obtained via the dimensional reduction.
In section \ref{sec4}, we apply our method to determine the tension from bulk EOM and boundary EOM to the O8-D8 brane systems. We can see that the correct value of the tension is reproduced. Section \ref{sec5} is for conclusion and comments on possible application to the replica manifold \cite{Lewkowycz:2013nqa} and connection to the cosmic brane \cite{Dong:2016fnf}.

\section{ETW branes from dimensional reduction}\label{sec2}

\subsection{Action and EOM}

In this section, we first explain our set-up and prepare the action and equations of motion (EOM) that will be used in later sections.

Our starting point is Einstein gravity theory with an $(n-1)$-form gauge field in $D=d+n+1$ dimensions. The action is
\be
S=\frac{1}{16\pi G_N^{(D)}}\int d^{D}x
\sqrt{-G_{(D)}}\left(R_{(D)}-\frac12 |f_n|_G^2-2\Lambda
\right),
\label{D_action}
\ee
where $G_N^{(D)}$ is the Newton's constant, $G_{(D)}=\det{(G_{AB}})$ is the determinant of the $D$-dimensional metric $G_{AB}$ ($A,B=0,\cdots,D-1$),
$R_{(D)}$ is the Ricci scalar curvature, $\Lambda$ is the cosmological constant, $f_n=\frac{1}{n!}f_{A_1\cdots A_n}dx^{A_1}\wedge\cdots\wedge dx^{A_n}$ is the field strength of the $(n-1)$-form gauge field and
\be
|f_n|_G^2=\frac{1}{n!}f_{A_1\cdots A_n}f_{B_1\cdots B_n}G^{A_1B_1}\cdots G^{A_nB_n}.
\ee

We consider a background parametrized by $(x^A)=(x^\mu,\theta^i,r,\tau)$ ($A=0,1,\cdots,D-1$; $\mu=0,1,\cdots,d-2$; $i=1,2,\cdots,n$), where $r$ is a radial coordinate bounded from below as $r\ge r_h$, $\tau$ is a periodic coordinate of period $\beta$, and $(\theta^i)$ are coordinates of the unit $n$-sphere $S^n$. The background configuration of the $n$-form field strength $f_n$ is assumed to be proportional to the volume-form of this $S^n$, and the metric is of the form
\be
ds_{(D)}^2=G_{AB}dx^Adx^B=W_1(r) (\eta_{\mu\nu}dx^\mu dx^\nu+f(r)d\tau^2)+W_2(r) dr^2+W_3(r)d\Omega_n^2 ,
\label{bg}
\ee
where $W_{1,2,3}(r)$ and $f(r)$ are non-negative functions of $r$, $\eta_{\mu\nu}$ is the $d-1$ dimensional Minkowski metric and $d\Omega_n^2$ is the line element of the unit $S^n$. The key assumption here is that the function $f(r)$ vanishes at $r=r_h$, while $W_{1,2,3}$ are non-zero there. The period $\beta$ is chosen so that the geometry is smooth everywhere including $r=r_h$. Therefore, the radius of the $S^1$ parametrized by $\tau$ shrinks to zero size at $r=r_h$ and the geometry contains a structure of the cigar depicted in the left panel of figure \ref{dimred}. This kind of geometry is realized as the (double) Wick rotation of a black $(d-1)$-brane solution, in which $\tau$ is the Euclidean time coordinate and $r$ is the radial coordinate transverse to the brane. The tip of the cigar at $r=r_h$ corresponds to the horizon of the black brane solution.

The next step is the dimensional reduction along the $\tau$ direction.\footnote{Here, we focus on the zero modes of the Fourier decomposition of the fields along the $\tau$ direction, assuming that the contribution from the non-zero modes can be neglected. In some examples considered in the next section, the radius of the $S^1$ diverges at the asymptotic boundary $r\rightarrow\infty$, where the dimensional reduction cannot be justified. These backgrounds can be regarded as the near horizon geometry of asymptotically flat backgrounds in which the radius of $S^1$ approaches a constant at $r\rightarrow\infty$. Since we are mainly interested in the physics around the tip of the cigar, we do not consider compactness seriously here.} The geometry after the dimensional reduction has a new co-dimension one boundary at $r=r_h$, which is regarded as an ETW brane.

Let us decompose the metric as
\be
ds_{(D)}^2=e^{-\alpha\Phi}\hat g_{MN}dx^Mdx^N+e^{(d+n-2)\alpha\Phi}d\tau^2,
\label{D_metric}
\ee
where $(x^M)=(x^\mu,\theta^i,r)$ ($M=0,1,\cdots,d+n-1$) are the $D-1=d+n$ dimensional coordinates, $\hat g_{MN}$ is the $D-1=d+n$ dimensional metric, $\Phi$ is the dilaton field and
\be
\alpha=\sqrt{\frac{2}{(d+n-1)(d+n-2)}}
=\sqrt{\frac{2}{(D-2)(D-3)}}\ .
\label{alpha}
\ee
Here, we have neglected the off-diagonal $(M,\tau)$ components of the metric. Similarly, we only keep the components of $f_n$ without the $\tau$ index. Then, the theory is reduced to the Einstein-dilaton graviy with $(n-1)$-form gauge field in a $D-1$ dimensional space-time with the ETW brane at $r=r_h$.
Inserting the metric (\ref{D_metric}) into the action (\ref{D_action})
the bulk action is obtained as
\begin{align}
S_{\rm bulk}=& \frac{1}{16\pi G_N}\int_{r>r_h} d^{d-1}x\,dr\,d^n\theta
\sqrt{-\hat{g}}
\left(\hat{R}-\frac{1}{2}\hat{g}^{MN}\p_M\Phi\p_N\Phi-v(\Phi)
%-\frac{1}{\sqrt{\hat{g}_{yy}}} t(\Phi)\delta(r_h)
-\frac{1}{2}w(\Phi)|f_n|^2\right)\ ,
\label{bulk_action}
\end{align}
where $G_N=G_N^{(D)}/\beta$, $\hat g=\det(\hat g_{MN})$, $\hat R$ is the Ricci scalar with respect to the $D-1$ dimensional metric $\hat g_{MN}$,
\be
|f_n|^2=\frac{1}{n!}f_{M_1\cdots M_n}f_{N_1\cdots N_n}\hat g^{M_1N_1}\cdots \hat g^{M_nN_n}
\ee
and
\be
v(\Phi)=2\,e^{-\alpha\Phi}\Lambda,~~~
w(\Phi)=e^{(n-1)\alpha\Phi}\ . 
\ee

In this paper, we analyze the ETW branes in two ways. One of them is to treat it as a boundary of the space-time, and add a boundary action that consists of Gibbons-Hawking term and the world-volume action with the dilaton dependent tension. This treatment will be discussed in section \ref{bdryEOM}. The other one is to regard the space-time as a ${\bf Z}_2$ orbifold with the ETW brane obtained as the fixed plane of the ${\bf Z}_2$ orbifold action. To be more explicit, we introduce a coordinate $y\in {\bf R}$ of the covering space related to $r$ as $|y|=r-r_h$, on which the ${\bf Z}_2$ acts as a sign flip $y\rightarrow -y$.\footnote{We actually take the orbifold fixed point to be slightly away from $r=r_h$, since $r=r_h$ is singular after the dimensional reduction. See section \ref{orbifold} for details.}
Then, working in the covering space, the ETW brane is a brane with dilaton dependent tension placed at $y=0$. The action for this system is
\begin{align}
S=& \frac{1}{16\pi G_N}\int d^{d-1}x\,dy\,d^n\theta
\sqrt{-\hat{g}}
\left(\hat{R}-\frac{1}{2}\hat{g}^{MN}\p_M\Phi\p_N\Phi-v(\Phi)
-\frac{1}{2}w(\Phi)|f_n|^2
-\frac{t(\Phi)}{\sqrt{\hat{g}_{yy}}}\delta(y)\right),
\nonumber\\
\label{action2}
\end{align}
where $t(\Phi)$ is the dilaton dependent tension of the ETW brane to be determined.

To obtain the background, we assume the following ansatz for the $D-1$ dimensional metric:
 \be
ds^2=\hat{g}_{MN}dx^Mdx^N=e^{2A(r)}g_{\mu\nu}(x)dx^\mu dx^\nu+e^{2B(r)}dr^2
+e^{2F(r)}d\Omega_n^2.
\label{ansatz}
\ee
The $d-1$ dimensional metric $g_{\mu\nu}(x)$, functions $A(r)$, $B(r)$, $F(r)$ and the dilaton field $\Phi(r)$ for the background (\ref{bg}) are given by
\begin{align}
&g_{\mu\nu}=\eta_{\mu\nu}\ ,\\
&e^{(D-3)\alpha\Phi(r)}=W_1(r)f(r)\ ,
\label{W1f}\\
&e^{2A(r)}=e^{\alpha\Phi(r)}W_1(r)\ ,\\
&e^{2B(r)}=e^{\alpha\Phi(r)}W_2(r)\ ,\\
&e^{2F(r)}=e^{\alpha\Phi(r)}W_3(r)\ .
\label{W3}
\end{align}
The $n$-form field strength $f_n$ is assumed to be of the form
\be
f_n=\frac{1}{n!}f_{i_1\cdots i_n}d\theta^{i_1}\wedge\cdots\wedge d\theta^{i_n}
=\frac{c}{n!}\ep_{i_1\cdots i_n}d\theta^{i_1}\wedge\cdots\wedge d\theta^{i_n},
\label{f_ansatz}
\ee
where $\theta^i$ ($i=1,2,\cdots,n$) are coordinates of the unit $S^n$ and $\ep_{i_1\cdots i_n}$ is the Levi-Civita symbol on $S^n$. Since the flux $\int_{S^n}f_n$ should be independent of $r$, the coefficient $c$ is a constant. Then, we obtain
\begin{align}
%    (f_n^2)_{ij}&=\frac{1}{(n-1)!}f_{ik_2\cdots k_n}f_j^{k_2\cdots k_n}=c^2e^{-2(n-1)F}g_{ij}^{(S^n)},\nn\\
    |f_n|^2&=\frac{1}{n!}f_{i_1\cdots i_n}f^{i_1\cdots i_n}=c^2e^{-2nF}.
\end{align}
Substituting these ansatz (\ref{ansatz}) and (\ref{f_ansatz}) into the action (\ref{action2}) and varying it with respect to $A$, $B$, $F$ and $\Phi$,
we obtain the following EOM:
\begin{align}
\frac{\delta S}{\delta A}=0~\Rightarrow~&
(1-2/(d-1)) e^{-2A+2B}R+n(n-1)e^{-2F+2B}
-2(d-2)(A''-A'B')
\nn\\
&
-2n(F''-F'B')
-(d-1)(d-2)(A')^2-2n(d-2)A'F'
-n(n+1)(F')^2
\nn\\
&=
\frac{1}{2} (\Phi')^2
+e^{2B}v(\Phi)
+\frac{c^2}{2} e^{-2nF+2B}w(\Phi)
+e^Bt(\Phi)\delta(y)\ ,
\label{EOMA}
\\
\frac{\delta S}{\delta B}=0~\Rightarrow~&
e^{-2A+2B}R+n(n-1)e^{-2F+2B}
\nn\\
&-\left((d-1)(d-2)(A')^2+n(n-1)(F')^2+2ndA'F'
\right)
+\frac{1}{2} (\Phi')^2
\nn\\
&=e^{2B}
\left( v(\Phi)+\frac{c^2}{2}e^{-2nF}w(\Phi)\right)\ , 
\label{EOMB}\\
\frac{\delta S}{\delta F}=0~\Rightarrow~&
 e^{-2A+2B}R+(n-1)(n-2)e^{-2F+2B}
-2(n-1)(F''-F'B')
\nn\\
&
-2(d-1)(A''-A'B')
-n(n-1)(F')^2-2(d-1)(n-1)A'F'
-d(d-1)(A')^2
\nn\\
&=
\frac{1}{2} (\Phi')^2
+e^{2B}v(\Phi)
-\frac{c^2}{2} e^{-2nF+2B}w(\Phi)
+e^Bt(\Phi)\delta(y)\ ,
\label{EOMF}
\\
\frac{\delta S}{\delta \Phi}=0~\Rightarrow~&
\Phi''+((d-1)A'+nF'-B') \Phi'
\nn\\
&=
e^{2B}\left(\frac{d v(\Phi)}{d\Phi}+\frac{c^2}{2}e^{-2nF}\frac{dw(\Phi)}{d\Phi}
\right)
+e^B\frac{dt(\Phi)}{d\Phi}\delta(y),
\label{EOMPhi}
\end{align}
where prime stands for the derivative with respect to $y$, e.g. $A'=\frac{\p A}{\p y}$, $A''=\frac{\p^2 A}{\p y^2}$, etc., and $R$ is the Ricci scalar for the $d-1$ dimensional metric. These EOM can also be obtained directly from the EOM derived from the action (\ref{action2}). Though we do not write down explicitly, the EOM for the metric $g_{\mu\nu}$ should also be imposed.

\subsection{Brane Tension from the bulk EOM via $\mathbf{Z}_2$ orbifold}
\label{orbifold}

The next task is to determine $t(\Phi)$ for given $A,B,F$ and $\Phi$. In this section we apply the $\mathbf{Z}_2$ orbifold method to the above bulk EOM (\ref{EOMA})--(\ref{EOMPhi}) to fix the tension. To do that, as mentioned in the previous section, we introduce the coordinate $y\in\mathbf{R}$ related to $r$ as
\be
r=\begin{cases}
    r_c+y,& \text{for \quad}y>0\ ,\\
    r_c-y,& \text{for \quad}y<0\ ,
\end{cases}
\ee
where $r_c$ is a cut-off satisfying $r_c>r_h$.
In this coordinate $y$, the geometry restricted to $r\ge r_c$ and its copy are glued together at $y=0$.
We will eventually take $r_c\rightarrow r_h$ limit. According to this, the functions $A(r),B(r)$, etc. in (\ref{ansatz}) can be redefined as a function of $y$ as
\be
A(y)
=\begin{cases}
A|_{r=r_c+y}, &\text{for \quad} y>0\ , \\
A|_{r=r_c-y}. &\text{for \quad} y<0\ .\\
\end{cases}
\ee
Taking the derivative with respect to $y$, we obtain
\be
A':=\p_y A(y)
=
\begin{cases}
\p_rA|_{r=r_c+y}, &\text{for}~ y>0 \ ,\\
-\p_rA|_{r=r_c-y}, &\text{for}~ y<0\ ,
\end{cases}
\ee
which leads to the discontinuity at $y=0$:
\be
 [A']^{+0}_{-0}=2\p_r A|_{r=r_c}\ .
\ee
Integrating the EOM (\ref{EOMA})-(\ref{EOMPhi}), we obtain
\begin{align}
&{[}-2(d-2)A'-2n F'{]}^{+0}_{-0}=e^{B(0)}t(\Phi(0)) \ ,
\label{E1}\\
&{[}-2(n-1)F'-2(d-1)A'{]}^{+0}_{-0}=e^{B(0)}t(\Phi(0))\ ,
\label{E2}\\
&{[}\Phi'{]}^{+0}_{-0}=e^{B(0)}\frac{dt(\Phi(0))}{d\Phi}\ ,
\label{E3}
\end{align}
where we pick up the contribution from the delta fucntion. Here, we have assumed $g'_{\mu\nu}$ is continuous at $y=0$.  From (\ref{E1}) and (\ref{E2}), we obtain
\begin{align}
&{[}A'{]}^{+0}_{-0}={[}F'{]}^{+0}_{-0}\ ,\\
&-2(d+n-2)[A']^{+0}_{-0}=e^{B(0)}t(\Phi(0))\ .
\label{E1-2}
\end{align}
When there is no sphere part (the case with $n=0$), the relevant EOM are (\ref{E3}) and (\ref{E1-2}).
(\ref{E3}) and (\ref{E1-2}) imply
\be
\frac{d}{d\Phi}\log t(\Phi)=-\frac{1}{2(d+n-2)}\frac{[\Phi']^{+0}_{-0}}{[A']^{+0}_{-0}}\ .
\label{E4}
\ee
The tension $t(\Phi)$ of the ETW brane can be determined by solving these equations as we will demonstrate in section \ref{sec3}.

\subsection{Brane Tension from the boundary EOM}
\label{bdryEOM}

It is also possible to determine the tension of the ETW brane by using the boundary EOM. Here, we explain how to do this and show that it leads to the same conditions as those found in the previous subsection.

Let us consider the space-time for $y>0$ with the ETW brane at $y=0$. The action consists of the bulk and boundary contributions as
\be
S=
% \frac{1}{16\pi G}\int_{y\ge 0} d^{d}xdy d^n\theta
% \sqrt{-\hat{g} }
%     \left(\hat{R}-\frac{1}{2}\hat{g}^{MN}\p_M\Phi\p_N\Phi-v(\Phi)
% -\frac{1}{2}w(\Phi)|f_n|^2
% \right)+
S_{\rm bulk}+ S_{\rm bdy}\ , 
\ee
where $S_{\rm bulk}$ is the bulk action given in (\ref{bulk_action}) integrated in the $y>0$ region. The boundary action $S_{\rm bdy}$ includes the Gibbons-Hawking term $S_{\rm GH}$ and the brane action  $S_{\rm brane}$ given by
\begin{align}
S_{\rm bdy}&=S_{\rm GH}+S_{\rm brane}\ ,\\
S_{\rm GH}&=\frac{1}{8\pi G_N}\int_{y=0} d^{d-1}x\,d^n\theta
\sqrt{-h}\, K\ ,
\\
S_{\rm brane}&=-\frac{1}{32\pi G_N}\int_{y=0} d^{d-1}x\,d^n\theta
\sqrt{-h}\,t(\Phi)\ ,
\label{braneaction}
\end{align}
where $h$ is the determinant of the induced metric $h_{ab}$ ($a,b=0,\cdots,d+n-2$) on the brane at $y=0$:
\be
 ds^2|_{y=0}=h_{ab}(x)dx^a dx^b
=e^{2A(0)}g_{\mu\nu}dx^\mu dx^\nu+e^{2F(0)}d\Omega_n^2\ ,
\ee
and $K=h^{ab}K_{ab}$ is the trace of the extrinsic curvature $K_{ab}$.
Note that the coefficient of $t(\Phi)$ in the brane action
(\ref{braneaction}) is a half of that in (\ref{action2}).
This is because (\ref{action2}) is the action for the covering space of the $\mathbf{Z}_2$ orbifold.\footnote{The tension $T$ in the original reference \cite{Takayanagi:2011zk} is related with the tension $t(\Phi)$ in (\ref{braneaction}) via
$T=\frac{t(\Phi)}{4}$.}

When the background metric is given by
\be
ds^2=dz^2+h_{ab}(x,z) dx^adx^b
\ee
(Gaussian normal coordinates), the extrinsic curvature
for the co-dimension one surface at $z=0$ is given by
\be
 K_{ab}=\frac{1}{2}\p_z h_{ab}|_{z=0}.
\ee
Using this formula, the extrinsic curvature for the background with metric  (\ref{ansatz}) can be computed as
\begin{eqnarray}
 K_{\mu\nu}&=&A'(0) e^{2A(0)-B(0)}g_{\mu\nu}(x)\ ,
 \label{Kmunu}\\
 K_{ij}&=&F'(0) e^{2F(0)-B(0)}g_{ij}(\theta)\ ,\\
K&=&g^{\mu\nu}K_{\mu\nu}+g^{ij}K_{ij}
=e^{-B(0)}(A'(0)d+F'(0)n)\ ,
\end{eqnarray}
where $g_{ij}(\theta)$ is the metric of the unit $S^n$:
\begin{eqnarray}
 d\Omega_n^2=g_{ij}d\theta^id\theta^j\ .
\end{eqnarray}

We impose the Neumann type boundary condition for the metric and the dilaton fields. The variation of the action with respect to the metric gives
the boundary terms as
\be
 \delta S=(\mbox{bulk terms})+\frac{1}{16\pi G_N}
\int_{y=0} d^{d-1}x\,d^n\theta\sqrt{-h}\,\delta h^{ab}
\left(
K_{ab}-K h_{ab}+8\pi G_N T_{ab}
\right)\ ,
\ee
where $T_{ab}$ is the energy momentum tensor for the brane
at the boundary
\be
 T_{ab}=\frac{2}{\sqrt{-h}}\frac{\delta S_{\rm brane}}{\delta h^{ab}}=-\frac{1}{32\pi G_N}t(\Phi)h_{ab}\ .
\ee
Therefore, the boundary equations of motion is
\be
 K_{ab}-K h_{ab}+8\pi G_N T_{ab}=0\ ,
\label{bEOM}
\ee
% In our case, we have
% \begin{align}
%  K_{\mu\nu}-K h_{\mu\nu}&=-(A'(d-1)+F' n)e^{2A-B}g_{\mu\nu}\ , \\
%  K_{ij}-K h_{ij}&=-(A'd+F'(n-1))e^{2F-B}g_{ij}\ , \\
%  T_{ab}&=-\frac{1}{32\pi G}t(\Phi)h_{ab}\ ,
% \end{align}
% where the right hand side of these equations is evaluated at $y=0$.
% Then, (\ref{bEOM}) implies
which implies
\begin{eqnarray}
 -4((d-2)A'+n F')|_{y=0}&=&e^{B(0)}t(\Phi)\ ,\\
 -4((d-1) A'+(n-1) F')|_{y=0}&=&e^{B(0)}t(\Phi)\ .
\end{eqnarray}
These equations are equivalent to (\ref{E1}) and (\ref{E2}).

Next, consider the variation with respect to $\Phi$.
The boundary terms for the variation of the action is
\be
 \delta S=(\mbox{bulk terms})+\frac{1}{16\pi G_N}
\int_{y=0} d^{d-1}x\,d^n\theta \sqrt{-h}\,\delta \Phi
\left(
\frac{1}{\sqrt{ g_{yy}}}\p_y\Phi
-\frac{1}{2} \frac{dt(\Phi)}{d\Phi}
\right),
\label{bdyvar}
\ee
which yields
\be
2\Phi'(0)= e^{B(0)}\frac{d t(\Phi(0))}{d\Phi}\ .\label{bdyEOM}
\ee
This agrees with (\ref{E3}).

\section{Examples}\label{sec3}

\subsection{AdS$_{d+1}$ soliton}\label{secsoliton}

Let us start with the case with $n=0$ and a negative cosmological constant parametrized as
\be
\Lambda=-\frac{d(d-1)}{2 L^2}\ ,
\label{cc}
\ee
for which the action (\ref{D_action}) is
\be
S=\frac{1}{16\pi G_N^{(d+1)}} \int d^{d+1}x\sqrt{-G}\left(R-\frac{d(d-1)}{L^2}\right)\ .\label{28}
\ee
We consider the $(d+1)$-dimensional AdS-soliton background \cite{Horowitz:1998ha,Constable:1999gb} given by
 \be
 ds^2=\frac{r^2}{L^2}\left(
 \eta_{\mu\nu}dx^\mu dx^\nu+f(r)d\tau^2\right)+\frac{L^2}{r^2}\frac{dr^2}{f(r)}\ ,
 \ee
 where $\mu,\nu=0,1,\cdots,d-2$ and
 \be
 f(r)=1-\frac{r_h^d}{r^d}\ .
 \ee
This is a solution of the EOM obtained from (\ref{28}).

In order to analyze the geometry near the tip $r=r_h$,
we introduce dimensionless coordinates $\tilde x^\mu$, $\tilde\tau$ and $z$ defined by
\be
{\tilde x}^\mu=\frac{r_h x^\mu}{L^2}\ ,~~~\tilde{\tau}=\frac{r_h\tau}{L^2}\ ,~~~
\frac{r^d}{r_h^d}=1+z^2\ .
\ee
For convenience, we omit the tilde of these coordinates and set $L=1$ in the following. Then, the metric becomes
\begin{align}
  ds^2&=G_{AB}dx^Adx^B\nn\\
  &=(1+z^2)^{\frac{2}{d}}\left(\eta_{\mu\nu}dx^\mu dx^\nu+\frac{z^2}{1+z^2}d\tau^2\right)+\frac{4dz^2}{d^2(1+z^2)}\label{27}\ ,
\end{align}
where $A,B=0,\cdots,d$. Then, the metric near $z=0$ becomes
\be
ds^2\simeq 
\eta_{\mu\nu}dx^\mu dx^\nu+z^2 d\tau^2+\frac{4}{d^2}dz^2\ ,
\ee
from which we find that the geometry is smooth at the tip $z=0$ if we choose the period of $\tau$ to be $\beta=4\pi/d$.

Using the ansatz (\ref{D_metric}) and (\ref{ansatz}), the metric is written as
\begin{align}
ds^2
&=e^{-\sqrt{\frac{2}{(d-1)(d-2)}}\Phi}\left(e^{2A}\eta_{\mu\nu} dx^\mu dx^\nu+e^{2B}dz^2\right)
+e^{\sqrt{\frac{2(d-2)}{d-1}}\Phi}d\tau^2\ ,
\label{solimetric}
\end{align}
with
\begin{align}
e^{2A}&=(1+z^2)^{\frac{1}{d}}z^{\frac{2}{d-2}}\ ,\nn\\
e^{2B}&=\frac{4}{d^2}(1+z^2)^{-\frac{1}{d}-1}z^{\frac{2}{d-2}}\ ,\nn\\
e^{\sqrt{\frac{2(d-2)}{d-1}}\Phi}&=z^2(1+z^2)^{\frac{2}{d}-1}\ .
\end{align}
Here, we have assumed $d>2$.

Following the procedure explained in section \ref{orbifold}, we introduce a coordinate $-\infty< y<\infty$ related to $z$ with $z=z_c+|y|$, where $z_c$ is a cutoff that will be sent to $z_c\rightarrow 0$ at the end. Then, near the orbifold fixed point $y=0$, $A$ and $\Phi$ behave as
\begin{align}
A&\simeq \frac{1}{d-2}\log(z_c+|y|)\ ,\nn\\
B&\simeq \frac{1}{d-2}\log(z_c+|y|)+\log(4/d^2)\ ,\nn\\
\Phi&\simeq\sqrt{\frac{2(d-1)}{d-2}}\log(z_c+|y|)\ ,
\end{align}
which lead to the discontinuity
\begin{align}
    [A']_{-0}^{+0}=\frac{1}{d-2}\frac{2}{z_c}\ ,~~~
    [\Phi']_{-0}^{+0}=\sqrt{\frac{2(d-1)}{d-2}}\frac{2}{z_c}\ .
\end{align}
Using this and (\ref{E4}), we obtain
\be
\frac{d}{d\Phi}\log t(\Phi)
=-\sqrt{\frac{d-1}{2(d-2)}}\quad\Rightarrow ~t\propto e^{-\sqrt{\frac{d-1}{2(d-2)}}\Phi}.\label{tension}
\ee
Substituting this into (\ref{E1-2}), we can also fix the coefficient as
\be
 t(\Phi)=-2d\,e^{-\sqrt{\frac{d-1}{2(d-2)}}\Phi}\ .
 \label{t}
\ee
Note that only for $d=3$ the coefficient of $\Phi$ in the exponent becomes integer.

Note that the tension (\ref{t}) diverges at $z=0$, since $\Phi\rightarrow-\infty$ as $z\rightarrow 0$. Indeed, the $d$-dimensional theory obtained via the dimensional reduction is singular at $z=0$. However, this singularity is resolved in the original $d+1$ dimensional theory. The situation is similar to a D6-brane in type IIA string theory and its M-theory lift. Note also that although the tension (\ref{t}) is negative, the system is stable, since the AdS-soliton background is believed to be stable.\cite{Horowitz:1998ha}

\subsection{AdS$_{d+1}$ soliton $\times$ S$^n$}

It is straightforward to generalize the discussion in the previous subsection to the case with $n>0$. The $D=d+n+1$ dimensional theory (\ref{D_action}) admits an AdS$_{d+1}$ soliton $\times S^n$ background with
\be
ds^2_{(D)}
=\frac{r^2}{L^2}
\left(\eta_{\mu\nu}dx^\mu dx^\nu+
f(r)d\tau^2
\right)
+\frac{L^2}{r^2}\frac{dr^2}{f(r)}+\tilde{L}^2d\Omega_n^2
\label{AdSbg}
\ee
with
\be
f(r)=1-\frac{r_h^d}{r^d}\ ,
\ee
where $\mu,\nu=0,1,\cdots,d-2$, and $L$, $\tilde L$ and $r_h$ are positive constants.
This metric takes the form (\ref{bg}) with
\be
W_1=\frac{r^2}{L^2}\ ,~~W_2=\frac{L^2}{r^2 f(r)}\ ,~~
W_3=\tilde{L}^2\ .
\ee
The configuration of the dilaton field $\Phi$ and the metric (\ref{ansatz}) for the $D-1=d+n$ dimensional theory obtained via dimensional reduction along the $\tau$ direction is obtained from the relations (\ref{W1f})--(\ref{W3}) as
\begin{align}
&e^{(D-3)\alpha\Phi(r)}=\frac{r^2}{L^2}f(r)\ ,
\label{Phi_D}\\
&e^{2A(r)}=\left(\frac{r^2}{L^2}f(r)\right)^{\frac{1}{D-3}}\frac{r^2}{L^2}\ ,\\
&e^{2B(r)}=\left(\frac{r^2}{L^2}f(r)\right)^{-\frac{D-4}{D-3}}\ ,\\
&e^{2F(r)}=\left(\frac{r^2}{L^2}f(r)\right)^{\frac{1}{D-3}}\tilde{L}^2\ ,
\label{F_D}
\end{align}
where $\alpha$ is defined in (\ref{alpha}).
Then, near the tip of the cigar $r=r_h$, $\Phi$ and $A$ behave as
\begin{align}
\Phi&\simeq \frac{1}{(D-3)\alpha}\log f(r)\ ,\\
A&\simeq\frac{1}{2(D-3)}\log f(r)\ ,
\end{align}
with which (\ref{E4}) implies
\be
\frac{d}{d\Phi}\log t(\Phi)
=-\frac{1}{(D-3)\alpha}
=-\sqrt{\frac{D-2}{2(D-3)}}\ ,
\ee
and
\be
t(\Phi)\propto e^{-\sqrt{\frac{D-2}{2(D-3)}}\Phi}\ .
\ee
Substituting this to (\ref{E1-2}), we obtain
\be
t(\Phi)=-2d\frac{r_h}{L^2}\,e^{-\sqrt{\frac{D-2}{2(D-3)}}\Phi}\ .
\label{tension_D}
\ee

\subsection{ETW branes in type 0A string theory}
In this section, we consider AdS soliton backgrounds (AdS$_4$-soliton $\times$S$^7$ and AdS$_7$-soliton $\times$S$^4$) in M-theory and take the dimensional reduction along the $\tau$ direction. Such backgrounds are obtained as supergravity solutions corresponding to M2 or M5-branes wrapped on a circle, along which the anti-periodic boundary condition is imposed on the fermions. This theory is conjectured to be dual to the type 0A string theory, because supersymmetry is completely broken by this boundary condition \cite{Bergman:1999km}. Correspondingly, the tip of the cigar in the AdS-soliton in M-theory is expected to become ETW brane in type 0A string theory. The tension of these objects can be calculated by using the results obtained in the previous subsection.

The low energy effective theory of M-theory is described by 11 dimensional supergravity, in which the massless bosonic fields are the metric and a 3-form gauge field. The action is given by (\ref{D_action}) with $D=11$, $n=4$, $d=6$ and $\Lambda=0$. The electromagnetic dual description obtained by taking the Hodge dual of the 4-form field strength is $D=11$, $n=7$ and $d=3$. It is known that it admits AdS$_4$-soliton $\times$S$^7$ and AdS$_7$-soliton $\times$S$^4$ backgrounds with (\ref{f_ansatz}) and (\ref{AdSbg}). In both of these backgrounds, (\ref{alpha}) implies $\alpha=1/6$ and (\ref{Phi_D})--(\ref{F_D}) become
\begin{align}
&e^{\Phi(r)}=\left(\frac{r^2}{L^2}f(r)\right)^{3/4}\ ,\\
&e^{2A(r)}=\left(\frac{r^2}{L^2}f(r)\right)^{\frac{1}{8}}\frac{r^2}{L^2}\ ,\\
&e^{2B(r)}=\left(\frac{r^2}{L^2}f(r)\right)^{-\frac{7}{8}}\ ,\\
&e^{2F(r)}=\left(\frac{r^2}{L^2}f(r)\right)^{\frac{1}{8}}\tilde{L}^2\ .
\end{align}
The tension (\ref{tension_D}) is
\be
t(\Phi)=-2d\frac{r_h}{L^2}e^{-\frac{3}{4}\Phi}\ ,
\ee
with $d=3$ and $d=6$ for AdS$_4$-soliton $\times$S$^7$ and AdS$_7$-soliton $\times$S$^4$ backgrounds, respectively.

Note that these results are for the Einstein frame. 
If we write the brane action (\ref{braneaction}) in the string frame as
\be
S_{\rm brane}=-\frac{1}{32\pi G_N} \int d^9 x\sqrt{-h_{\rm st}}\, t_{\rm st}(\Phi)
\ ,
\ee
where $h_{\rm st}$ is the determinant of the induced metric in the string frame ($h^{\rm st}_{ab}=e^{-\Phi/2}h_{ab}$), the tension is
\be
t_{\rm st}(\Phi)=-2d\frac{r_h}{L^2}e^{-3\Phi}\ .
\label{tst}
\ee

The peculiar dilaton dependence in (\ref{tst}) suggests that this ETW brane is a non-perturbative object in type 0A string theory. In fact, in perturbative string theory, the scattering amplitudes depend on the dilaton as $e^{-\chi\Phi}$, where $\chi$ is the Euler characteristic of the string world-sheet. Since the Euler characteristic $\chi$ is given by $\chi=2-2g-b-c$, where $g$, $b$ and $c$ are the number of handles, boundaries and crosscaps of the world-sheet, respectively, $\chi$ is at most $2$ and it is not possible to realize $t(\Phi)\sim e^{-3\Phi}$ found in (\ref{tst}) in the perturbative expansion. This is not a contradiction, because our calculation is based on M-theory description, which corresponds to the strongly coupled regime in string theory. Note that the dilaton dependence $e^{-3\Phi}$ has been observed, for example,
in the tension of the S-dual of a D7-brane in type IIB string theory.

\section{The O8-D8 brane system}\label{sec4}

In section \ref{sec2}, we explained how to determine the tension of the ETW brane via bulk EOM and boundary EOM. In this section, as a consistency check, we apply these methods to a system with an O8-plane\footnote{Here, we consider a negative tension orientifold 8-plane called O8$^-$-plane.} and D8-branes in type IIA string theory, and show that it reproduces the correct value of the tension of these objects.

Let us consider type IIA string theory in ${\bf R}^9\times I$, where $I=[0,\ell]$ is an interval parametrized by $u$, with an O8-plane at each boundary $u=0,\ell$ and 16 parallel D8-branes localized along $I$.\footnote{This system is obtained as the T-dual of type I string theory and is also called as type I' or type IA string theory.}
We place $N$ D8-branes at $u=0$ and $16-N$ D8-branes at $u=\ell$. The corresponding  supergravity solution \cite{Horowitz:1991cd,Polchinski:1995df,Bergshoeff:1996ui} in the Einstein frame reads
\begin{align}
 ds_{\rm E}^2&=g_{MN}dx^M dx^N,\nn\\
&=H(u)^{1/8}
\eta_{\mu\nu} dx^\mu dx^\nu
+H(u)^{9/8}du^2\nn\\
&=:
e^{2A}\eta_{\mu\nu} dx^\mu dx^\nu
+e^{2B}du^2\ ,\label{metric}
\end{align}
($M,N=0,1,\cdots,9$; $\mu,\nu=0,1,\cdots,8$) with
\begin{align}
 e^{2A}&=H(u)^{1/8}\ ,\\
 \label{A_O8}
 e^{2B}&=H(u)^{9/8}\ ,\\
 H(u)&=a+\frac{g_s (8-N)}{2\pi l_s}u\ ,
 \label{H_O8}
\end{align}
and
\be
e^\Phi=g_s H(u)^{-\frac{5}{4}}\ ,
\label{Phi_O8}
\ee
where  $g_s$ is the string coupling constant, $l_s$ is the string length, and $a$ is a constant. Below, we set $g_s=1$ for simplicity.

The interval $I$ can be regarded as $S^1/{\bf Z_2}$, where ${\bf Z_2}$ is the orientifold action. Here, $S^1$
is parametrized by $u\in [-\ell,+\ell]$, where $u=-\ell$ and $u=+\ell$ are identified, and the ${\bf Z_2}$ acts on $u$ as $u\rightarrow -u$. In this case, $u$ in (\ref{H_O8}) is replaced with $|u|$ and hence the derivatives of $A$, $B$ and $\Phi$ with respective to $u$ have discontinuity at $u=0,\ell$. Then, substituting (\ref{A_O8}) and (\ref{Phi_O8}) into (\ref{E1-2}) and (\ref{E4}), the tension $t(\Phi)$ for the ETW brane at $u=0$ is obtained as
\be
t(\Phi)=\frac{N-8}{\pi l_s}e^{\frac{5}{4}\Phi}\ .
\label{tO8D8}
\ee
This is the dilaton dependent tension of the O8-plane with $N$ D8-branes at $u=0$ in the Einstein frame appeared in (\ref{braneaction}):
\be
 S_{\text{O8-D8}}=-\frac{1}{32\pi G_N}\int d^{9}x\,du\sqrt{-h}\, t(\Phi)\ ,\label{actiond8}  
\ee
where the Newton constant $G_N$ is given by
\be
G_N=\frac{(2\pi l_s)^8}{32\pi^2}\ .
\ee
Then, the action (\ref{actiond8}) reproduces the correct tension of the O8-plane and D8-branes in the string frame:
\begin{align}
    S_{\text{O8-D8}}&=-\frac{N-8}{(2\pi)^8l_s^9}\int d^9x\, e^{-\Phi}\sqrt{-h_{\rm st}}\nn\\
    &=-(N-8)\,T_{\rm D8}\int d^9x\, e^{-\Phi}\sqrt{-h_{\rm st}},
\end{align}
where $T_{\rm D8}=1/((2\pi)^8l_s^9)$ is the tension of the D8-brane and $h_{\rm st}$ is the determinant of the induced metric on the ETW brane at $u=0$ in the string frame.

 We can also confirm that the boundary equations of motion are satisfied. The dilaton EOM obtained from (\ref{bdyvar}) is satisfied as
\be
\frac{\p_u \Phi}{\sqrt{g_{uu}}}=-\frac{5}{4}H'H^{-\frac{25}{16}}=
\frac{N-8}{2\pi l_s}\frac{5}{4}e^{\frac{5}{4}\Phi}=\frac{1}{2}\frac{dt}{d\Phi}\ .
\ee
Similarly, using the explicit expression for the extrinsic curvature
\be
K_{\mu\nu}=\frac{1}{16}H'H^{-\frac{23}{16}}\eta_{\mu\nu}=
\frac{8-N}{32\pi l_s}H^{-\frac{23}{16}}\eta_{\mu\nu}\ ,
\ee
obtained from (\ref{Kmunu}),
the boundary EOM for the metric (\ref{bEOM}) can be checked as 
\be
K_{\mu\nu}-Kh_{\mu\nu}=-\frac{8-N}{4\pi l_s}H^{-\frac{23}{16}}\eta_{\mu\nu}
=-\frac{8-N}{4\pi l_s}e^{\frac{5}{4}\Phi}e^{2A}\eta_{\mu\nu}
=-\frac{1}{4}t(\Phi)h_{\mu\nu}
=-8\pi G_N T_{\mu\nu}
\ .
\ee

\section{Conclusion and Discussion}\label{sec5}
In this paper, we have considered the dimensional reduction of the cigar geometries, where Euclidean time direction shrinks at the tip. The key observation was that the tip of the cigar became an ETW brane after the dimensional reduction. To derive the tension of the ETW brane, we used a trick to regard the ETW brane as the fixed plane on a $\mathbf{Z}_2$ orbifold. We also gave another derivation of the tension using boundary EOM, which leads to the same results. As a check of our calculation, we applied our methods to the O8-D8 brane system and confirmed that
the known value of the tension of the O8-plane with $N$ D8-branes is reproduced consistently.

In particular, we considered the AdS$_7$-soliton $\times$S$^4$ and the AdS$_4$-soliton $\times$S$^7$ backgrounds in M-theory and took the dimensional reduction along the Euclidean time direction $\tau$, which is regarded as the M-theory direction $x^{11}$. This theory is conjectured to be type 0A string theory \cite{Bergman:1999km} with the ETW brane corresponding to the tip of the cigar of the AdS-soliton geometry. We calculated the tension of the ETW brane by the above methods and found that the tension is proporthional to $t(\Phi)\sim e^{-3\Phi}$ in the string frame in both cases. This dilaton dependence does not appear in a perturbative expansion series with respect to the string coupling constant, and hence it should be considered as a non-perturbative object in string theory.

Although we mainly focused on the AdS-soliton geometries in the examples considered in section \ref{sec3}, our methods can be applied to other cigar geometries. One interesting geometry with the cigar structure is the Euclidean Schwarzschild black hole geometry where $\tau$ is the Euclidean time direction and $r_h$ is the horizon. It would be extremely interesting if the ETW brane obtained via the dimensional reduction play some role in the black hole thermodynamics.

As another possible application of the dimensional reduction of the cigar geometry, let us make a few comments on a replica manifold in gravitational path integral \cite{Lewkowycz:2013nqa} which have been used for the derivation of the Ryu-Takayanagi formula in AdS/CFT \cite{Ryu:2006bv,Ryu:2006ef}. This is an Euclidean path integral and the replicated manifold can be described by the cigar geometry. In general, the $n$-replicated manifold is related via holographic dual to the R\'{e}nyi-$n$ entropy $S_n$ in CFT. This geometry is described by the cigar geometry with a cosmic brane with tension $T_n=\frac{n-1}{4n G}$ \cite{Dong:2016fnf} at the fixed point of the replica symmetry. At this stage, our set up and this geometry is quite different since in this case, it contains a co-dimension two cosmic brane even in the background geometry (even before the dimensional reduction). However, we can take the $n\rightarrow0$ limit and the replica manifold has a boundary at the position of the fixed point of the $\mathbf{Z}_N$ symmetry. In the gravity side, the co-dimension two cosmic brane reduces to the co-dimension one ETW brane and the tension of the brane is fixed as $T_0=-\lim_{n\to 0}\frac{1}{4nG}$, which has infinitely negative value as in our previous results. Although we naively analytically continued the tension $n\rightarrow0$ here, we expect that this tension can be reproduced as in our explicit calculation of dimensional reduction especially in AdS$_3$ case. For the CFT side, $n\rightarrow0$ limit of the R\'{e}nyi entropy is called Hartly entropy $S_0=\log |\mathcal{A}|$, where $\mathcal{A}$ is the dimension of the Hilbert space of the subsystem (See \cite{Agon:2023tdi} for a recent work on field theory analysis). This is dual to the area of the cosmic brane and especially ETW brane for $n\rightarrow0$. Therefore, we do not expect the tension of the brane itself is related with the dimension of the Hilbert space, but it will be interesting to study in more detail along this direction.

\section*{Acknowledgement}
We thank T.Kawamoto, T.Takayanagi and S.Terashima for stimulating discussions. We also thank the participants of the 18th Kavli Asian Winter School, where a part of the work was presented.
The work of SS is supported by JSPS KAKENHI (Grant-in-Aid for Scientific Research (B)) grant number JP19H01897 and MEXT KAKENHI (Grant-in-Aid for Transformative Research Areas A "Extreme Universe") grant number 21H05187. YS is supported by Grant-in-Aid for JSPS Fellows No.23KJ1337.

\bibliographystyle{JHEP}
\bibliography{soli.bib}

\end{document}